# LYMAN LIMIT IMAGING OF HIGH-REDSHIFT GALAXIES. III. NEW OBSERVATIONS OF 4 QSO FIELDS

CHARLES C. STEIDEL[1,2,3]
MIT, Physics Department, Room 6-201, Cambridge, MA 02139
e-mail: ccs@astro.caltech.edu

MAX PETTINI
Royal Greenwich Observatory, Madingley Road, Cambridge CB3 OEZ, UK
e-mail: pettini@ast.cam.ac.uk

DONALD HAMILTON
Max Planck Institute fur Astronomie, Konigstuhl 17, D–69117 Heidelberg, GERMANY
e-mail: hamilton@mpia-hd.mpg.de

## ABSTRACT

We present the results of deep Lyman limit imaging in four new fields as part of a continuing search for galaxies at $3.0 \leq z \leq 3.5$. The technique uses a custom broad–band filter set ($U_nG\mathcal{R}$) designed to isolate objects having Lyman continuum breaks superposed on otherwise flat–spectrum ultraviolet continua. The observations are specifically aimed at detecting known galaxies producing optically thick QSO metal line absorption systems, but are equally sensitive to more generally distributed objects at the redshifts of interest.

We have identified a plausible candidate for the QSO absorber in one out of the four new fields surveyed; in the other three the absorbing galaxy must be either fainter than $\mathcal{R} = 25.5$ or closer than $\sim 1$ arcsec to the QSO sight-line (or both). Together with previously published results of the survey, we have now detected in two out of six cases objects with the expected properties of star–forming galaxies at $z \gtrsim 3$ within 3 arcsec of the QSO sight lines; the two candidate absorbers have similar luminosities, $M_B \simeq -22$ ($q_0 = 0.5$, $H_0 = 50$ km s$^{-1}$Mpc$^{-1}$), and impact parameters, $R \sim 10\ h^{-1}$ kpc.

We find the average surface density of robust Lyman break objects to be $\sim 0.5$ galaxies per square arcminute to a magnitude limit $\mathcal{R} = 25.0$ . A simple, "no evolution" model based on the properties of normal galaxies at $z \lesssim 1$ predicts a density of Lyman break objects only a few times larger than observed. We conclude that there is a substantial population of star–forming galaxies, of relatively normal luminosity, already in place at $z = 3 - 3.5$. If normal galaxies experienced a period of very high star formation early on in their history, it must have occurred prior to $z = 3.5$.

---

[1]Alfred P. Sloan Foundation Fellow

[2]NSF Young Investigator

[3]Current address: Palomar Observatory, Caltech 105-24, Pasadena, CA 91125

It is possible that the observed surface density of $z > 3$ objects is biased by the presence of a luminous QSO in the redshift range of interest; the surveyed fields are too small to examine the possibility of a spatial association of the $z > 3$ candidates with the QSOs. We suggest a number of future observations that would resolve this question and clarify the nature of the putative "normal" galaxy population beyond $z \sim 3$.

## 1. INTRODUCTION

One of the major themes of current observational cosmology is the search for high redshift galaxies. Unless we can identify and study field galaxies from the earliest times to the present day, our understanding of galaxy formation and evolution will remain largely theoretical speculation. Of particular interest is the epoch $z \gtrsim 3$ which some considerations suggest may be the time when luminous galaxies first assembled. For $H_0 = 50$ km s$^{-1}$ Mpc$^{-1}$ and $q_0 = 0.05$ (adopted throughout this paper), this redshift corresponds to a look-back time of $\sim 14$ Gyr, within the range of ages of Galactic globular clusters (Chaboyer 1995). Studies of metal line systems in QSO absorption spectra appear to indicate a significant increase of the heavy-element content of the universe from $z \simeq 3$ (Steidel 1990; Pettini et al. 1995a), accompanied by a decrease in the comoving density of neutral gas consistent with consumption by star formation (Wolfe et al. 1996). This is also the epoch of galaxy formation in models based on a cold dark matter universe (e.g. Haehnelt & Rees 1993; Kauffmann & Charlot 1994).

Searches for galaxies at $z \gtrsim 3$ have been spectacularly *un*successful up to now, given the efforts devoted to the quest. The two main techniques employed have each come across difficulties which were largely unforeseen at the outset. Deep imaging at optical wavelengths has revealed a population of faint blue galaxies which are mostly at $z \lesssim 1$ (Colless 1995 and references therein; Guhathakurta, Tyson & Majewski 1990). Since these galaxies dominate the counts at the faintest magnitudes, their presence complicates significantly the identification of objects at higher redshifts. Surveys for Lyman $\alpha$ emission, both in blank areas of sky and in the fields of known QSO absorbers, have similarly produced generally null results (Djorgovski, Thompson, & Smith 1993; Pettini et al. 1995b; Lowenthal et al. 1995). With the benefit of hindsight we now appreciate the many reasons—mostly to do with the high optical depths encountered by Lyman $\alpha$ photons—why Lyman $\alpha$ emission is not a prominent spectral feature in most astrophysical environments (Charlot & Fall 1993; Neufeld 1991; Chen & Neufeld 1994).

In this paper, the third in a series, we continue to explore the feasibility of a third approach: ultra-deep imaging near the Lyman limit. While H II regions have no strong emission lines in the ultraviolet (e.g. Hartmann et al. 1988; Terlevich et al. 1993; Deharveng, Buat, & Bergeron 1995), we expect the Lyman break at 912 Å to be a ubiquitous feature in galaxy spectra (e.g.

Leitherer & Heckman 1995). Even massive stars emit relatively few photons below the Lyman limit; this intrinsic discontinuity in the integrated spectrum of a young stellar population is likely to be further enhanced by absorption by interstellar H I within the star-forming, and therefore presumably gas-rich, galaxy. As it turns out, however, our method is not particularly sensitive to any assumptions about the intrinsic spectral energy distribution or the self–absorption in the vicinity of 912 Å, as the combined, statistical effects of known sources of opacity due to intervening gas are guaranteed to produce an effective Lyman continuum decrement in the far–UV spectrum of any object in the targeted redshift range (see §3).

The situation is illustrated in Figure 1 where we show the spectral energy distribution (SED) of a generic star-forming galaxy at $z = 3.151$, as predicted by the spectral synthesis models of Bruzual & Charlot (1993) which assume a Salpeter IMF. The spectrum is characterized by a blue continuum and an intrinsic drop by a factor of $\sim 5$ across the Lyman jump; the only other spectral feature which can be discerned is the strong stellar Lyman $\alpha$ absorption line. It would therefore appear that the most promising strategy in the search for such galaxies is very deep imaging through filters specifically chosen to isolate the two most obvious features, the Lyman break and the blue continuum, as shown in Figure 1. The red $U_n - G$ and blue $G - \mathcal{R}$ colors of a galaxy at $z = 3.151$ should readily differentiate it from other objects in the field.

This technique can be applied to blank fields of sky. We have however chosen to begin our survey in directions where we already know of the existence of a $z > 3$ galaxy from the Lyman limit absorption it produces in the spectrum of a background QSO. Apart from the obvious rationale of looking first where one is most likely to find the object sought, we are motivated by the particularly powerful combination of QSO spectroscopy on one hand, and deep imaging on the other, for assessing the evolutionary status of high redshift galaxies. Nevertheless, as we shall see, our results are not limited to the QSO sight-lines.

Pilot observations of the field of the QSO Q0000−2620 ($z_{\rm em} = 4.11$), reported by Steidel & Hamilton (1992, 1993; Papers I and II respectively), demonstrated the effectiveness of Lyman limit imaging by isolating 16 candidates for galaxies near $z = 3.390$, the redshift of the Lyman limit system (and one of the two damped Lyman $\alpha$ systems) in the QSO spectrum. One of the galaxies is close to the QSO sight-line ($10 - 19\ h^{-1}$ kpc, where $h$ is the Hubble constant in units of 100 km s$^{-1}$ Mpc$^{-1}$) and is probably one of the two damped Lyman $\alpha$ absorbers[4]; another shows Lyman $\alpha$ emission at $z_{\rm em} = 3.428$ (Giavalisco, Macchetto, & Sparks 1994; Giavalisco et al. 1995). An analysis of the angular distribution of the 16 candidates on the plane of the sky suggested that many of the objects are likely to be part of a group or cluster of galaxies (Giavalisco, Steidel, & Szalay 1994).

In this paper we present the results for the fields of another 4 QSOs with Lyman limit systems at $z_{\rm abs} > 3$. Although the overall sample is still very small, the new cases allow us to make a better assessment of the feasibility of the method and to draw some preliminary conclusions on the luminosity of high-redshift galaxies.

[4]At the time of writing Paper I, we were aware of only the $z_{abs} = 3.390$ damped system; another (lower column density) LLS/damped system is present at $z_{abs} = 3.050$ (Savaglio, D'Odorico, & Moller 1994). Only one candidate $z > 3$ galaxy was found near the QSO sightline.

## 2. OBSERVATIONS AND REDUCTIONS

### 2.1. Filter Bandpasses

All the images were obtained with the same set of custom–made filters used in the observations of Q0000−2620 and designated $U_n$, $G$, and $\mathcal{R}$. The spectral responses of these filters are reproduced in Figure 1 and have been discussed in detail in Paper II. The measured counts through the filters were placed directly onto an AB magnitude system (Oke & Gunn 1983) by comparison with spectrophotometric standard stars from the compilations by Stone & Baldwin (1983) and Massey et al. (1988); internal errors in the photometric zero points are smaller than 0.02 magnitudes in all 3 passbands. The transformations from the $U_nG\mathcal{R}$ system onto the standard Johnson/Kron-Cousins filter system are given in Paper II.

### 2.2. Data Acquisition

In this paper we bring together observations of four QSOs obtained at three different observatories in the period from 1992 September to 1994 April; relevant details are collected in Table 1.

Images of the field of Q0347−3819 were obtained in 1992 November at the Cassegrain focus of the 4 m telescope of the Cerro Tololo Inter-American Observatory. A thinned Tektronix $2048 \times 2048$ pixel CCD, binned $2 \times 2$, gave a scale of $0\farcs32$ per (binned) pixel and a field size of $4.3 \times 4.3$ arcminutes after accounting for a small amount of vignetting produced by the 2-inch square filters. The skies were clear but the seeing was mediocre, ranging from 1.5 to 2.0 $''$ FWHM.

Data for the other three QSO fields were obtained with the 4.2 William Herschel telescope at La Palma, Canary Islands; Q2233+1310 was observed in 1992 September, and Q1244+1129 and Q1451+1223 in 1993 April. We used a thinned Tektronix $1024 \times 1024$ pixel CCD at the Cassegrain auxiliary port; this combination provides high UV throughput and fine spatial sampling ($0.109''$ per unbinned $24\mu$m pixel), but results in a very small field of view, which is roughly circular with $\sim 100$ arcsecond diameter (the field is vignetted at the corners). The observing conditions during both WHT runs were not the most favorable, and limited the total exposure time which could be achieved for each object. While the seeing was generally good for the useful portions of the runs (particularly during 1993 April, where it was typically 0.75-0.85 $''$ ), the transparency was at times less than optimum. For this reason we used the 2.4 m Hiltner telescope at the Michigan–Dartmouth–MIT Observatory, under clear conditions and in good seeing, to obtain supplemental data in the $G$ and $\mathcal{R}$ bands for Q2233+1310 (1993 October) and in the $\mathcal{R}$ band for Q1451+1223 (1994 April). The detector was a thinned Tektronix 1024 $\times$ 1024 pixel CCD, providing a scale of 0.275 $''$ per pixel.

We used a standard in–field dithering technique during all of the observations, with typical individual exposures of 600 s in $\mathcal{R}$, 900 s in $G$, and 2000 s in $U_n$. The long exposure in $U_n$ ensured that the images are sky-noise (rather than read–out noise) limited, even with the small pixel size of the WHT observations. We were careful to include in each image at least one star, brighter than the QSO, from which the PSF could be modeled accurately. Given the small field of the WHT observations, in one case (Q1451+1223) this necessitated placing the QSO near the edge of the detector.

### 2.3. Data Reduction

The CCD images from all of the observing runs were reduced in a similar fashion, using the procedures described in detail in Paper II. The individual exposures were put into sub-pixel co-registration and added; in cases where more than one telescope was used to obtain data for the same field, all of the images were re-sampled to the WHT pixel scale, translated, and rotated using the IRAF "geotrans" task (preserving flux). In the case of Q2233+1310, the $\mathcal{R}$ WHT and MDM images were co-added after this process, weighting by the total number of electrons collected for objects in the field. The total exposure times and the resulting image quality and depth for the final frames are listed in Table 1.

The galaxies causing the Lyman limit absorption in the QSO spectra are likely to be located within a few arc seconds of the QSO line of sight; therefore we took a great deal of care in subtracting the QSO light profile as accurately as possible by modeling the PSF of bright stars in the field. Despite our efforts, the subtraction process at times still left significant residuals within approximately half of the PSF FWHM from the QSO position. Presumably the PSF changes by small, yet significant, amounts over the area of the CCD, possibly due to imperfections in the telescope optics, the filters, or the flatness of the detector. In any case, this limitation of the technique makes it difficult to detect with certainty objects closer to the QSO than $\sim 0.5''$ (closer than $\sim 1''$ in the case of Q0347$-$3819 – see Table 1). On the other hand, we found that at angular separations greater than the PSF FWHM there is no significant additional (Poisson) noise introduced by the subtraction process, and the detection limit is essentially the same as that at any random position in the image.

### 2.4. Photometry

Faint galaxy photometry was performed with FOCAS (Valdes 1982), adopting a procedure which has been discussed in detail in Paper II. Briefly, we defined an initial sample with a conservative cutoff in $\mathcal{R}$ magnitude, $\mathcal{R} = 25.5$. The justification for this relatively bright limit stems from the expectation that the high-redshift objects sought will be somewhat fainter in $G$ than in $\mathcal{R}$, and significantly fainter in $U_n$ than in $G$. In our deep images $\mathcal{R} = 25.5$ is a highly significant detection ($\sim 10 - 15\sigma$; see Table 1). For an object to be included in the initial catalog we required that, after convolution with the standard FOCAS smoothing kernel, the number of adjacent pixels exceeding 3 times the local sky $\sigma$ corresponds to an area greater than that subtended by the FWHM of the seeing disk. In practice, the average isophotal size of an object with $\mathcal{R} = 25.5$ is more than 3 times the area encircled by the FWHM of the seeing profile. The $\mathcal{R}$ isophotal apertures were applied directly to the $G$ and $U_n$ images; in this way each object in the frames was measured through an optimized "aperture" defined by the light profile in the $\mathcal{R}$ band.

Both isophotal and FOCAS "total" $\mathcal{R}$ magnitudes of each object were retained; the difference between the isophotal and total magnitudes in the $\mathcal{R}$ frame was then used as an aperture correction for each object, and applied to the measurement in *each* bandpass (this assumes that there is no significant radial color gradient for the faint galaxies). The average correction varied from $\sim 0.02$ magnitudes at $\mathcal{R} = 22$ to $\sim 0.10$ magnitudes at $\mathcal{R} = 25.5$. Colors were obtained from the difference in measured isophotal magnitudes; to minimize errors, we smoothed (when necessary) all three images of a given QSO field to the FWHM of the image with the worst seeing.

We made no attempt to discriminate between stars and galaxies; while we are confident that this can be done to relatively faint magnitudes given the good seeing for all but the Q0347−3819 field, we do not want to exclude compact galaxies near the detection limit by misclassifying them as stars. Most of the faint stars in our deep images are likely to be M stars which are clearly separated from the locus of galaxies in color-color plots.

Objects for which the flux in the $G$ band was less than $2\sigma$, where $\sigma$ is the effective sky noise inside the isophotal aperture, were excluded for the purposes of measuring colors. Such red objects will not yield useful constraints on the $U_n - G$ color in any case. Consequently, in the final catalog used in this paper, some of the reddest galaxies near the limit $\mathcal{R} = 25.5$ are excluded. For objects which are detected in both $G$ and $\mathcal{R}$, but with $U_n$ band fluxes of less than 1 $\sigma$ (inside the isophotal aperture), the $U_n$ magnitude was flagged as a limit (at the 1 $\sigma$ value). Thus, the limits for galaxies undetected in the $U_n$ band depend on their isophotal size in the $\mathcal{R}$ band.

In the last column of Table 1 we give estimates of foreground reddening due to Galactic dust in the four directions studied. The values listed are based on H I column density measurements from the survey by Stark *et al.* (1992) and the average $< N(\mathrm{H}^0)/\mathrm{E(B{-}V)} >= 5.27 \times 10^{21}$ cm$^{-2}$ mag$^{-1}$ found by Diplas & Savage (1994). We also estimate that E$(U_n - G)$ = 1.25 E$(B - V)$ and E$(G - \mathcal{R})$ = 1.26 E$(B - V)$. Therefore, Galactic extinction is unlikely to have a measurable effect on the colors measured in 3 out of our 4 fields, and may introduce a marginal reddening ($\sim 0.1$ mag) in the Q2233+1310 field in both $U_n - G$ and $G - \mathcal{R}$. We have made no corrections to the measured magnitudes in any of the fields.

## 3. THE COLORS OF HIGH-REDSHIFT GALAXIES

In this section we consider quantitatively the color criteria applied in the selection of candidate high-redshift galaxies. Figure 2 shows the expected locus of galaxies in the $(U_n - G)$ vs. $(G - \mathcal{R})$ plane as a function of redshift and spectroscopic type. In calculating the galaxy colors we have combined the "generic" E, Sb, Sc, and Im spectral energy distributions by Bruzual & Charlot (1993) with Madau's (1995) statistical estimate of the opacity introduced by line blanketing in the Lyman alpha forest and by intervening Lyman limit systems. We have *not* included—because its magnitude is unknown although the effect is most likely present— self-absorption in the Lyman continuum by interstellar gas *within* the galaxies. Even relatively small column densities of neutral gas will increase significantly the decrement below the Lyman limit ($N(\mathrm{H}^0) = 1.5 \times 10^{17}$ cm$^{-2}$ will produce an additional decrement by one magnitude); consequently the $U_n - G$ colors plotted in Figure 2 are probably underestimates at redshifts $z > 3.3$, where the $U_n$ bandpass falls completely shortward of the Lyman break.

A consequence of our choice of the AB magnitude system and of the effective wavelengths of the three filters (§2.1) is that a galaxy with a smooth spectral energy distribution of the form $f_\nu \propto \nu^{-\alpha}$ will have *equal* $U_n - G$ and $G - \mathcal{R}$ colors in the absence of line blanketing; a galaxy with a flat SED ($\alpha = 0$) will lie at $(U_n - G) = (G - \mathcal{R}) = 0$ in the diagram.

The most striking feature in Figure 2 is the very rapid increase in the $U_n - G$ color for $z \gtrsim 2.8$, accompanied by an appreciable, albeit less pronounced, reddening in $G - \mathcal{R}$. In contrast, the typical galaxy colors at lower redshifts are $(G - \mathcal{R}) \approx (U_n - G) \approx 0.3 - 0.8$, corresponding to a spectral energy distribution $f_\nu \propto \nu^{-\alpha}$ with $0.7 \leq \alpha \leq 2.0$. Note that beyond $z \approx 2.8$ the UV color is essentially independent of the galaxy spectroscopic type. The reason for this is that the major contributor to the increasingly red $U_n - G$ color with redshift is Lyman continuum absorption caused by intervening material *external* to the galaxy itself; the intrinsic Lyman break of the input stellar energy distributions calculated by Bruzual & Charlot (1993) is typically only $\sim 1.5 - 2$ magnitudes. *The point here is that the overall picture of the UV color evolution shown in Figure 2 is based more on known sources of opacity—the Lyman $\alpha$ forest and Lyman limit systems (Madau 1995)—than our assumptions about the galaxies SEDs*[5]

---

[5]We should point out that the blanketing due to extrinsic absorption is a statistical calculation, and that fluctuations are possible along a particular line of sight; because of this, and because of photometric errors, reddening, and the details of the assumptions going into the model SEDs, the expected spectral energy distributions of the model galaxies should be taken as a rough guide rather than as precise predictions.

The dotted lines (and the dashed line at $G - \mathcal{R} = 0$) in Figure 2 encompass the region in the color–color plane where we expect to find galaxies with $3.0 \lesssim z \lesssim 3.5$ in our images. The diagonal dotted line corresponds to colors such that the flux decrement between $U_n$ and $G$ is more than 4 times that between $G$ and $\mathcal{R}$. The vertical line at $(G - \mathcal{R}) = 1.2$ arises because in galaxies at $z > 3.5$ the $G$ bandpass is sufficiently absorbed that the observed colors become indistinguishable, despite the large $U_n - G$ breaks, from those of intrinsically red galaxies at more modest redshifts.[6] This is readily realized when we consider that the detection limits achieved in the $U_n$ bandpass are between $U_n \simeq 27.0$ and 27.4 (see Table 1). Consequently, at the faintest magnitudes in our $\mathcal{R}$−selected catalogue of objects, $\mathcal{R} \leq 25.5$, we only have a $\sim 2$ magnitude "dynamic range" for the detection of the Lyman break. At these faint magnitudes we must rely on only the bluest objects in $G - \mathcal{R}$ to constrain our search.

In addition to the redshift dependence of the expected $G - \mathcal{R}$ colors, there is also a redshift dependence of the expected "detectability" of a galaxy in the $U_n$ band. This dependence, which has been explored comprehensively by Madau (1995), has mostly to due with the quantity of flux longward of the Lyman continuum break that "leaks" into the $U_n$ filter, which is fixed in the observed frame. There will be such "leaks" to varying degrees until $z \approx 3.3$, at which point the $U_n$ bandpass is completely in the Lyman continuum. The measured $U_n - G$ color of the QSO in each of our fields (columns 5 and 6 of Table 2) provides a rough empirical estimate of the color we can expect for objects with $z = z_{LLS}$ (particularly when $z_{abs} \approx z_{em}$), since the QSOs were chosen to have no measurable flux shortward of the Lyman limit.

Figure 2 shows that over the small frequency baseline sampled by our photometry, only a spectral break is capable of introducing enough apparent curvature to give colors that stand out obviously in the color–color diagram. Apart from the Lyman limit, the only other such discontinuity is the Balmer continuum/4000 Å break. This will not affect the $U_n - G$ color (apart from galaxies at zero redshift!) but will be straddled by the $G$ and $\mathcal{R}$ filters for $0.2 \leq z \leq 0.7$. Galaxies in this redshift range with both intermediate–age stars and current star formation, may exhibit a $G - \mathcal{R}$ color that is as red as $\sim 3$ magnitudes, yet have a $U_n - G$ color corresponding to a nearly flat spectrum (see Figure 2). This is simply due to the fact that the *shape* of the far-UV continuum of galaxies is dependent almost exclusively on the current rate of star formation and the behavior of the stellar initial mass function at the high mass end.

To conclude, we have searched each of the four QSO fields for faint objects which lie within the dotted lines in Figure 2, these being the most likely candidates for high-redshift galaxies. In order to reduce misidentifications which may result from photometric scatter, edge effects, and similar problems (such as faint galaxies paired with bright ones, which can lead to incorrect measurements in the "splitting" process attempted by FOCAS), we subsequently examined each candidate on the final reduced images. The objects which passed this vetting procedure are believed to be free of obvious photometric problems. We now briefly discuss each QSO field in turn.

---

[6]The line blanketing due to the Lyman $\alpha$ forest becomes quite severe beyond $z \sim 3.5$, so that even if one were to devise a program to straddle the Lyman continuum break at a specific redshift using closely–spaced intermediate band filters (e.g, de Robertis and McCall 1995), the sensitivity for the detection of breaks would be greatly reduced, and the broad–band colors would be difficult or impossible to distinguish from intrinsically red field galaxies.

## 4. DISCUSSION OF INDIVIDUAL FIELDS

### 4.1. Q0347−3819

This QSO, discovered by Osmer & Smith (1977), has been studied at low dispersion by Lanzetta et al. (1991), and at higher resolution by Williger et al. (1989) and Steidel (1990). The known metal line systems are summarized in Table 3. The Lyman limit system at $z_{\rm abs} = 3.025$ is also a damped Lyman $\alpha$ absorber, with log $N$(H I) = 20.7 and a metallicity of less than 1/6 of solar (Pettini et al. 1994).

This field has both the largest area ($3'.82 \times 3'.82$ in the final summed image) and the poorest seeing ($\approx 2''$ ) of the four fields studied; however, the long integration times in $U_n$ and $G$ and the dark sky at CTIO resulted in images which are nearly as deep as those of the other three fields for the present purpose (see Table 1). In Figure 3 we have reproduced contour plots of $80'' \times 80''$ portions of the $\mathcal{R}$, $G$, and $U_n$ images centered on the QSO position. Table 4 gives the results of our photometry for all objects within $60''$ of the QSO sight line; objects which are not detected in $U_n$ are listed separately at the end of the Table and are indicated by the prefix 'N'.

Figure 4 is the color–color plot for all objects in the *full* field satisfying our selection criteria ($\mathcal{R} \leq 25.5$ and $\geq 2\sigma$ significance in $G$; §2.4). Compared to the other fields observed, there appears to be an unusually high surface density of relatively bright ($\mathcal{R} \lesssim 23$) galaxies in the direction of Q0347−3819. Note that most objects do indeed occupy the locus predicted for galaxies at redshifts ($z \lesssim 2$); the very red objects, with $(G - \mathcal{R}) \simeq 2$ and $(U_n - G) \gtrsim 2$, have the colors expected for late-type stars and indeed are all relatively bright.

We find no obvious candidate for the LLS/DLA absorber near the QSO sight-line. The closest object, number 1 in Figures 3 and 4 and in Table 4, has a separation of $3''.1$ from the QSO and colors $(G - \mathcal{R}) = 0.80$, $(U_n - G) = -0.11$. These are the colors expected for a galaxy at the redshift of the strong Mg II $z_{\rm abs} = 1.4571$ system (see Figure 2); in this case, the galaxy would have $M_B \approx -22.0$, or $\approx 2.5\ L*$ (adopting a $k$ correction appropriate for an Sc galaxy spectroscopic type), and a separation $\approx 18\ h^{-1}$ kpc from the QSO. Both values are within the range encompassed by Mg II absorbers at $z \simeq 0.6$ (Steidel, Dickinson, & Persson 1995); we therefore consider G1 as the prime candidate for the $z_{\rm abs} = 1.4571$ system. Object number 2, $8''.4$ west of the QSO, is too red in $G - \mathcal{R}$ to qualify as a $z \sim 3$ candidate under our criteria, and is also probably too distant from the QSO sight-line to be responsible for the damped Lyman $\alpha$ system.

The third object within $10''$ of the QSO, labeled 'G' in Figure 3, has $\mathcal{R} = 25.2$. This galaxy is not reliably detected in the $G$ frame, but can just be discerned in the $U_n$ band image ($U_n = 25.9$) and is therefore unlikely to be at $z \sim 3$. Thus, it appears that the $z_{\rm abs} = 3.0244$ Lyman limit absorber is either fainter than $\mathcal{R} = 25.5$ or closer than $2''$ from the QSO (or both).

On the other hand, as can be seen from Figure 4, there are 8 objects (out of a total of 398 in the $3'.82 \times 3'.82$ field) with the colors of $z > 3$ galaxies. Of these the most noteworthy is N5, located $26''$ north and $1''.5$ west of the QSO (see Figure 3). Its $U_n - G$ color is more extreme than that of the QSO, suggesting that it may be at a *higher* redshift; at $z = 3.3$ (the redshift at which the $U_n$ bandpass is completely below the Lyman limit), the observed $\mathcal{R} = 23.82$ would imply $M_B \approx -23.9$, or $\approx 15\ L*$ (applying a $k$ correction appropriate for an Im galaxy at these redshifts). The resolution of our images is insufficient to assess whether N5 is stellar (in which case it may be a low-luminosity AGN) or extended; recent NTT observations at $0.9''$ resolution, however, indicate that it is indeed extended (Giavalisco 1995, private communication). Thus N5 appears to be a luminous high-redshift galaxy and as such deserves further study; the object is sufficiently 'bright' to be accessible to low-resolution spectroscopy.

### 4.2. Q1244+1129

This QSO is from the "Large Bright Quasar Survey" of Foltz et al. (1987). To our knowledge there are no subsequent published observations at higher resolution than the discovery spectrum; we therefore used periods of poor seeing during the 1993 April run to secure $\sim 3.5$ Å FWHM resolution spectra of the QSO with the WHT Cassegrain spectrograph and EEV CCDs (Figure 5). Table 5 lists the metal lines detected, including Lyman $\alpha$ lines in identified metal systems. Lines numbers 1 and 2 have been flagged by Lanzetta et al. (1991) as possible damped Lyman $\alpha$ lines. ¿From their widths we estimate log $N$(H I)$= 20.1 \pm 0.1$ and log $N$(H I)$= 20.3 \pm 0.1$ at $z_{\rm abs} = 2.6347$ and $z_{\rm abs} = 3.0950$ respectively, but higher resolution data are required to assess the degree of line blending and confirm these values. Some blending is almost certainly present, as indicated by the asymmetric profiles, the systematically lower redshifts of both Lyman $\alpha$ lines compared to metal lines in the same absorption systems (see Table 5), and the generally high density of the Lyman $\alpha$ forest at these wavelengths. The Lyman limit break due to the higher redshift system occurs near 3737 Å (beyond the short-wavelength limit of the spectrum in Figure 5) as noted by Sargent & Steidel (1989 unpublished; see Stengler-Larrea et al. 1995) and by Lanzetta et al. (1991). As can be seen from Figure 5 and Table 2, this absorption system is only $\sim 3000$ km s$^{-1}$ to the blue of the QSO emission redshift.

Images of the Q1244+1129 field are reproduced in Figure 6; the results of the photometry are collected in Table 6 and plotted in the color-color diagram in Figure 7. Although there are several very faint objects within $10''$ of the QSO, every one of them is detected in all three passbands, thereby failing our selection criteria for a $z > 3$ candidate. Some of these objects are too faint to be included in Table 6, although they are clearly real because of the good positional match in the three images. Presumably the galaxies responsible for the lower redshift systems along this sight-line, at $z_{\rm abs} = 2.636$, 1.951 and 1.2833, are among these faint objects. In particular, object number 3 ($\mathcal{R} = 23.62$, $7''$ west of the QSO) is the leading candidate for the $z_{\rm abs} = 1.2833$ Mg II absorber; if the identification is correct, the galaxy has $M_B \approx -22.2$ (applying a $k$ correction appropriate to an Sc galaxy spectroscopic type) and a halo extending at least 40 $h^{-1}$ kpc (to cover the QSO sight-line).

As can be seen from Figure 7, there are 3 – 4 candidates for $z > 3$ galaxies in this field. We draw attention in particular to G8, $12\farcs5$ south of the QSO, with $\mathcal{R} = 24.59$, $(G - \mathcal{R}) = 0.08$ and $(U_n - G) = 2.66$. This object is marginally detected in $U_n$ and has colors very similar to those of the QSO. If at $z = 3.1$, the galaxy has $M_B \approx -23.0$ (applying the $k$ correction of an Im galaxy spectroscopic type at these redshifts); its separation from the QSO is 78 $h^{-1}$ kpc.

### 4.3. Q1451+1223

We could find no spectra of this faint QSO, discovered by Hazard et al. (1986), of sufficient resolution and S/N to carry out a comprehensive census of absorption systems. Low–resolution spectra of relatively low S/N have been obtained by Lanzetta et al. (1991) and by Sargent & Steidel (1989, unpublished); the latter spectrum is reproduced in Figure 8. The Lyman limit system at $z_{\rm abs} = 3.171$ is identified on the basis of the break near 3802 Å and the Lyman $\alpha$ line at 5070.6 Å. Lanzetta et al. also noted two candidate damped Lyman $\alpha$ systems at $z_{\rm abs} = 2.254$ and 2.470, but better data are required to confirm their reality and to identify other metal line systems which are likely be present in the spectrum, given the high redshift of the QSO ($z_{\rm em} = 3.247$).

Contour plots of the images of the Q1451+1223 field are presented in Figure 9, the color–color diagram in Figure 10, and the photometric results in Table 7. As in the previous two cases, the Lyman limit absorber is apparently below our detection limit, although we do find high-$z$ candidates further away from the QSO. Of the four objects closest to the QSO, numbers 2 and 3 are probably stars as they are unresolved in $0\farcs83$ seeing, while 1 and 4 have the colors of typical field galaxies at these faint magnitudes. With a better QSO spectrum it should be possible to check for the presence of metal-line systems produced by galaxies 1 and 4. Of the eight high-$z$ candidates in Figure 10, five have $U_n - G$ colors which are $\approx 2$ mag bluer than that of the QSO, suggesting that they may be at lower redshifts than the Lyman limit system. Of the other three N1 is the most luminous; the measured $\mathcal{R} = 24.5$ corresponds to $M_B \approx -23.2$ at $z = 3.2$.

### 4.4. Q2233+1310

The QSO was discovered by Crampton, Schade, & Cowley (1985). An intermediate dispersion spectrum was published by Sargent, Steidel, & Boksenberg (1989) who identified the Lyman limit system at $z_{\rm abs} = 3.151$ and suggested the presence of additional metal systems at $z_{\rm abs} = 2.828$, 2.556, 2.492, and 1.026. Lu et al. (1993) deduced $N$(H I)= $1 \times 10^{20}$ cm$^{-2}$ from profile fitting of the Lyman $\alpha$ line associated with the Lyman limit system, and confirmed the reality of the $z_{\rm abs} = 2.556$ system. In Figure 11 we show new spectra covering the wavelength range 3200 – 9000 Å, obtained with the Kast double spectrograph of the Shane 3.0 m telescope at Lick Observatory. The spectral resolution is $\sim 4.5$ Å (comparable to that of the Sargent et al. data) in the regions 3200 – 5400 and 6600 – 9000 Å, and $\sim 2.3$ Å between 5400 and 6600 Å. Table 8 lists the metal lines identified. The new data do *not* confirm the reality of the $z_{\rm abs} = 2.828$, 2.492, and 1.026 systems, but reveal a previously unrecognized C IV doublet at $z_{\rm abs} = 2.660$.

Figures 12 and 13 show the contour maps and color–color plot of the Q2233+1310 field; the photometry is collected in Table 9. In this case we do have a strong candidate for the Lyman limit absorber in N1, only 2.9″ from the QSO. The object is clearly not an artifact of the PSF subtraction, as it is seen at the same position in both the WHT and MDM $\mathcal{R}$ images (as well as the MDM $G$ image). Placed at $z = 3.151$ this galaxy would have $M_B \approx -22.6$, or $\approx 4\ L*$, and dimensions exceeding 18 $h^{-1}$ kpc (to intercept the QSO sight-line). These values are similar to those deduced in Paper I for the Lyman limit absorber in front of Q0000−2620 ($M_B \approx -23.1$, adopting the same $k$ correction as here, and projected radius greater than 19 $h^{-1}$ kpc). With $(G - \mathcal{R}) = 0.55$ and $(U_n - G) > 2.08$, N4 may well be another high-$z$ candidate in this field.

## 5. DISCUSSION

Together with the earlier observations of Q0000−2620 (Papers I and II), we have imaged the fields of five QSOs with six Lyman limit systems to approximately similar levels of sensitivity for detecting galaxies at $z \gtrsim 3$. In two cases, Q0000−2620 and Q2233+1310, we have identified plausible candidates for the absorbing galaxies; the two candidates have similar luminosities, colors and impact parameters from the QSO sight-lines. The other four Lyman limit absorbers are either fainter than an apparent magnitude $\mathcal{R} = 25.5$ or closer than $\sim 1$ arcsecond from the QSO sight-line (or both).

In all five fields we have detected a number of other faint objects, at projected distances too large to be the absorbers, but nevertheless with the colors expected for galaxies at $z > 3$. Pending spectroscopic confirmation, we have constructed a subset of “robust” candidates from the objects which fall within the area bounded by the dotted lines in the color-color diagrams in Figures 4, 7, 10 and 13. We consider a candidate “robust” if it is either *undetected* in $U_n$ or is detected and has the same colors as the QSO; that is, we exclude objects which are detected in the $U_n$ band even though they occur near the locus occupied by high-$z$ galaxies. In addition, given the limited depth in the $U_n$ band and our stringent criteria for defining $z > 3$ candidates, we consider only objects having $\mathcal{R} \leq 25.0$ as "robust" (fainter objects having $G - \mathcal{R}$ colors redder than the middle of the expected range would have insufficient limits on $U_n - G$ to qualify as a robust candidate). We have adopted this conservative approach because the lower bound of the region corresponding to $z > 3$ in the two-color plot is somewhat arbitrary; on the other hand the "robust" sub–sample almost certainly underestimates the total number of $z > 3$ objects in our images. For example, galaxy G2 in the field of Q0000−2620, with Lyman *alpha* emission at $z = 3.428$ (Giavalisco et al. 1995), is *not* included because it lies just below the expected region of the color-color plot using our adopted (more stringent) criteria.[7]

Table 10 summarizes the results for the five fields. Within the uncertainties resulting from the small number statistics (an inevitable consequence of the limited field of view of the WHT observations), we find the same average surface density of high-$z$ candidates in all five directions. From the weighted mean of the five determinations we conclude that the density of galaxies with $\mathcal{R} \leq 25.0$ in the interval $z = 3 - 3.5$ (this being the redshift range to which we are sensitive—see Figure 2) is $\approx 0.5$ arcmin$^{-2}$. We now show that this detection rate is approximately as expected if there has been little evolution in the population of galaxies which give rise to QSO Lyman limit systems from $z \lesssim 1$ to $z \gtrsim 3$.

We make our model predictions of the surface density galaxies in the range $3.0 \leq z \leq 3.5$ by taking advantage of the fact that the redshift path density of QSO Lyman limit absorption systems is now well–established to $z \sim 4$, with a co–moving total cross-section consistent with no evolution (Sargent *et al.* 1989), and the connection between these absorption systems and the overall field galaxy population has now been established at moderately high redshift. The extensive survey of Mg II absorbing galaxies (at a given redshift Mg II and Lyman limit systems trace the same population of absorbers) by Steidel, Dickinson, & Persson (1994, 1995) has shown that galaxies selected by absorption cross-section at an average redshift $\langle z \rangle = 0.65$ (and redshift range $0.3 \leq z \leq 1.0$) provide an essentially complete census of relatively luminous ($L_K > 0.1L_K^*$), massive galaxies, with a luminosity function similar to that of present-day Hubble sequence field galaxies. A self-consistent approach to testing the "no evolution" hypothesis is therefore to base our calculation of the expected rate of incidence of high-$z$ galaxies on the well-established properties of the Mg II (LLS) absorbers at $z \lesssim 1$ and the *observed* incidence of the absorption systems at the appropriate redshifts. In this way, we assume only that the slowly evolving gaseous envelopes are associated with the same type of object from $z \sim 3$ to $z \lesssim 1$[8] (In fact, preliminary

---

[7]However, subsequent deeper $U_n$ images obtained at the ESO NTT place G2 well within the expected region of the diagram (Giavalisco *et al.* 1995, private communication). Only about half of the candidates discussed by Giavalisco *et al.* 1994 satisfy our "robust" criteria, although most of these candidates are confirmed, with greater significance, by the new NTT data.

[8]Since the luminosity function of the absorbing galaxy population is so similar to that of samples selected by apparent magnitude, particularly brighter than $L^*$, estimates of the galaxy surface density based on no evolution

observations of the absorption–selected galaxies in the redshift range $1.0 \leq z \leq 1.6$ suggest little or no evolution in space density, gaseous size, or K–band luminosity relative to the $\langle z \rangle = 0.65$ sample [Steidel & Dickinson 1995]).

Our estimate of the expected number of $z > 3$ galaxies is based on three considerations, as follows:

1. Figure 14 shows that our "robust" limit $\mathcal{R} \leq 25.0$ samples only the bright end of the luminosity function; depending on the value of $q_0$, we are only sensitive to galaxies brighter than $\approx 2.8L_B^*$ ($q_0 = 0.05$), or $\approx 1.0L_B^*$ ($q_0 = 0.5$). From the luminosity distribution of Mg II absorbers determined by Steidel et al. (1994) we find that for $q_0 = 0.05$, 7% of the galaxies are brighter than $2.8L_B^*$, while for $q_0 = 0.5$, 25% are brighter $1.0L_B^*$. (Obviously the unknown $k$ correction from observed $\mathcal{R}$, corresponding to $\sim 1750$ Å at $z = 3$, to rest–frame $B$ is highly relevant here. The values in Figure 14 are based on the assumption that the UV/blue spectrum of a typical $z = 3$ galaxy resembles that of a generic "Im" galaxy in the spectral synthesis code of Bruzual & Charlot (1993). Since the major contributors at UV/blue wavelengths are short-lived, massive stars, the spectral shape is probably independent of age to a first approximation; the IMF and the metallicity also have only minor effects on the UV slope [see Figures 31–34 of Leitherer & Heckman 1995]. Thus, provided the $z > 3$ galaxies detected in our images are actively forming stars, the $k$ correction we have adopted is likely to be approximately correct.)

2. We assume that the typical cross–section for a Lyman limit system is $\pi R^2$, where $R$ is the average radius of the spherical halos producing Mg II absorption at $< z >= 0.65$. From Steidel (1995), $R = 38\ h^{-1}$ and $33.5\ h^{-1}$ kpc for $q_0 = 0.05$ and 0.5 respectively. At $z = 3.25$ (the middle of the redshift range to which we are sensitive) 1 arcsecond corresponds to $6.1\ h^{-1}$ and $3.5\ h^{-1}$ kpc for $q_0 = 0.05$ and 0.5 respectively. Thus, if the gaseous sizes of galaxies have not changed appreciably between $z \sim 1$ and 3.25, we expect the typical angular cross-section of a LLS halo at $z = 3.25$ to be 120 or 290 arcsec$^2$ for $q_0 = 0.05$ and 0.5 respectively.

3. The redshift path density of LLS, $dN/dz = 2 \pm 0.5$ at $z \simeq 3$, measured by Sargent et al. (1989) and Stengler-Larrea et al. (1995), implies that over the redshift interval $z = 3 - 3.5$ every line of sight intercepts one Lyman limit absorber on average.

From points 2 and 3 it follows that the surface density of absorbing galaxies, *of all luminosities*, with $z = 3 - 3.5$ is 1 per 120 arcsec$^2$ (or 30 arcmin$^{-2}$) for $q_0 = 0.05$, and 1 per 290 arcsec$^2$ (or 12 arcmin$^{-2}$) for $q_0 = 0.5$. Since for $q_0 = 0.05$ we are only sensitive to the brightest ~7% of galaxies in the luminosity function (point 1), we conclude that under the assumption of "no evolution" we expect a surface density of 2.1 arcmin$^{-2}$; the corresponding density for $q_0 = 0.5$ (where under our assumptions we can observe the brightest ~25% of the galaxy luminosity

---

of the local galaxy luminosity function, or a moderate redshift galaxy sample selected by apparent magnitude (e.g., Lilly *et al.* 1995), will result in similar predictions; we simply regard it as a slightly smaller "leap of faith" to connect moderate redshifts to high redshifts through the absorption line systems, which are well-studied over the entire range.

distribution) is 3.1 $\mathrm{arcmin}^{-2}$ .

These values are $\approx 4-6$ times *higher* than the observed surface density of "robust candidates", $\approx 0.5$ $\mathrm{arcmin}^{-2}$ which, as argued above, probably underestimates the true number of high-$z$ galaxies. Inspection of Figures 4, 7, 10 and 13 suggests that the underestimate could be as large as a factor of $\sim 2$, based on the numbers of objects having $[(G-\mathcal{R})\leq 1.2]$ and marginally significant lower limits to $(U_n - G)$. Thus, estimated and observed numbers of high-$z$ galaxies are in reasonably good agreement, given the rough nature of our calculation and the large correction factors which have to be applied to account for the fraction of the luminosity function which is below the detection limit of the present data. The crucial point, however, is that we do *not* detect significantly *more* galaxies than expected under the null assumption of "no evolution". Nevertheless, our results, while still preliminary, suggest that a population of galaxies with ultraviolet luminosities and gaseous dimensions roughly similar to moderate redshift galaxies in the Hubble sequence, was already in place at $z \gtrsim 3$ . Consistent with the null results of searches for "primeval" galaxies, we do not find evidence for extremely high star formation rates at $z = 3-3.5$ .

It could be argued that, by centering our searches on luminous QSOs, we may be *overestimating* the average density of high-$z$ galaxies in the field. This would be the case if many of our candidates turned out to be at the same redshifts as the QSOs (possible for all but Q0000−2620, which has a much higher emission redshift than the range to which our technique is sensitive). This is certainly a possibility, given that five candidates in the field of Q0347−383 and one in the field of Q1244+1129 have the same colors as the respective QSOs. Our images cover too small an area of sky to address this question by examining the spatial distribution of objects relative to the QSOs. While from Figure 15 it can be seen that the $z > 3$ candidates in the field of Q0347−3819 are not significantly clustered near the QSO, the transverse dimensions of even this field, the largest in the present work, are only 1.4 $h^{-1}$ Mpc ($q_0 = 0.05$; 0.8 $h^{-1}$ Mpc for $q_0 = 0.5$), barely beyond the core radius of present-day rich clusters of galaxies.

## 6. CONCLUSIONS AND FURTHER WORK

We have analyzed deep $U_nG\mathcal{R}$ images of the fields of four QSOs with Lyman limit systems at $z > 3$ and reached two main conclusions:

1. The new data confirm initial indications (Papers I and II) that the technique of Lyman limit imaging is one of the most effective strategies for recognizing galaxies at $z \gtrsim 3$ .

2. Although the sample is still very small, the observed surface density of high-redshift galaxies is roughly comparable with the value expected on the basis of known properties of absorption–selected galaxies at $z \lesssim 1$ . Apparently, the UV luminosity and gaseous extent of what would become Hubble sequence galaxies have not changed radically from $z \simeq 3$ to the present time. The presence of a substantial population of apparently "normal", luminous galaxies at $z \gtrsim 3$ may present challenges to many hierarchical models of structure formation. If normal galaxies experienced a period of very high star formation early on in their history, it must have occurred prior to $z = 3.5$.

Several important follow-up observations are suggested by the present results. Most importantly at present, spectroscopic confirmation of the $z > 3$ candidates is required, in at least some of the fields, to establish what fraction of the faint objects with no flux in the $U_n$ band are indeed high-$z$ galaxies. This difficult task may now be within reach with 8-10 m class telescopes.

Extending the photometry (and, possibly, the spectroscopy) to the near-infrared would provide information on the rest-frame optical properties of the galaxies and clues to the evolutionary status of their stellar populations and a more robust estimate of the mass function of the young, star–forming galaxies.

Deeper $U_n$ frames would permit a more reliable separation of Lyman limit objects from the general locus of faint blue galaxies at lower redshifts. The most significant improvement in the determination of the surface density of $z > 3$ galaxies, however, will result from increasing the sensitivity of surveys such as this one *in all three passbands*, so as to sample a greater proportion of the luminosity function. If the luminosity function of the $z \lesssim 1$ Mg II/LLS absorbers is representative, pushing the limits reached here one magnitude deeper should result in 5–2.5 times more candidates ($q_0 = 0.05$ and 0.5 respectively).

Finally, it would obviously be valuable to extend this work to random areas of sky (and, in order to improve the statistics on the relatively rare candidates, to obtain data in much larger fields), so as to ascertain whether the surface density of galaxies is higher than normal in the fields of bright QSOs and thereby assess how common large-scale structures were at $z > 3$. It would clearly be most advantageous to coordinate this effort with deep HST imaging which can provide a glimpse of the morphology of these galaxies even at such high redshifts.

We are grateful to the CTIO, MDM and WHT time assignment committees for their generous support of this demanding observational program. We also thank Mauro Giavalisco and Piero Madau for many conversations and for communicating results in advance of publication. CCS acknowledges the support of grant AST-9457446 from the NSF.

---

# FIGURE CAPTIONS

Fig. 1.—Spectral energy distribution of a star-forming galaxy at $z = 3.151$ from the population synthesis models by Bruzual & Charlot (1993); the example reproduced here is for a generic Im galaxy and assumes a Salpeter IMF. The broken lines show the transmission curves of the three broad-band filters used in this work, chosen specifically to detect the Lyman break at 912 Å ($U_n$ and $G$) and the relatively flat continuum longward of the break ($G$ and $\mathcal{R}$). Also shown in the figure is the spectrum of the $z_{\rm em} = 3.295$ QSO Q2233+1310; an optically thick Lyman limit system at $z_{\rm abs} = 3.151$ produces the marked discontinuity near 3900 Å. Details of the QSO spectrum are given in §4.4 .

Fig. 2.—Color evolution of galaxies of different spectroscopic type in the three passbands used in this work; points are plotted at redshift intervals $\Delta z = 0.1$ starting from $z = 0.0$. In producing the plot we have combined the spectral energy distributions by Bruzual & Charlot (1993) with Madau's (1995) statistical estimates of Lyman line and continuum blanketing by intervening gas. No allowance has been made for Lyman absorption by the interstellar medium of the galaxies themselves. The dotted line indicates the locus of points which we expect to be occupied by high-redshift galaxies ($z \gtrsim 3$).

Fig. 3.—Contour plots of the $80''$ region surrounding the QSO Q0347−3819. The lowest contour corresponds to a surface brightness of $\sim 1$ sigma above sky after lightly smoothing the images; contour spacing is logarithmic with spacing of 0.12 in log(surface brightness). Objects discussed in the text are marked. The profile of the QSO has been subtracted from each image; its position is marked by a cross at the center of the coordinate system.

Fig. 4.—Two-color diagram for the field of Q0347−3819, including all objects with $\mathcal{R} < 25.5$. Open circles are objects detected in all 3 passbands; the size of the circle scales inversely with $\mathcal{R}$ magnitude. Open triangles indicate objects detected in both $\mathcal{R}$ and $G$, but not in $U_n$. The QSO itself is represented with a heavy "X". As discussed in the text, the dotted lines enclose the region of the diagram where galaxies with $3.0 \leq z \leq 3.5$ are expected to lie.

Fig. 5.—Spectrum of Q1244+1129 obtained with the ISIS spectrograph on the WHT 4.2 m telescope; the resolution is $\sim 3.5$ Å FWHM. Significant metal absorption lines are listed in Table 5. The features near 7600 Å are residuals from an imperfect correction of the atmospheric "A" band.

Fig. 6.—Same as Figure 3, for the Q1244+1129 field.

Fig. 7.—Same as Figure 4, for the Q1244+1129 field.

Fig. 8.—Low resolution (4–6 Å FWHM) spectrum of Q1451+1223 obtained with the Double Spectrograph on the Hale 5.08 m telescope.

Fig. 9.—Same as Figure 3, for the Q1451+1223 field. The QSO position is near the bottom of the images, for reasons discussed in the text.

Fig. 10.—Same as Figure 4, for the Q1451+1223 field.

Fig. 11.—Spectrum of Q2233+1310 obtained with the Kast Spectrograph on the Shane 3.0 m telescope at Lick Observatory; the resolution is $\sim 4.5$ Å FWHM for the first and third panels, and 2.3 Å FWHM for the second panel. Significant metal absorption lines are listed in Table 8.

Fig. 12.—Same as Fig. 3, for the Q2233+1310 field.

Fig. 13.—Same as Fig. 4, for the Q2233+1310 field. Note that the QSO falls at the very top of the diagram, with $(U_n - G) = 4.1$ and $(G - \mathcal{R}) = 0.4$ .

Fig. 14.—Plot showing the expected apparent $\mathcal{R}$ magnitude at the redshifts of interest here of a galaxy with the blue luminosity of a present–day $L^*_B$ galaxy. For our assumed cosmology ($q_0 = 0.05, H_0 = 50$ km s$^{-1}$ Mpc$^{-1}$) an $L^*_B$ galaxy has $M^*_B = -21.0$. We have applied the $k$ correction appropriate to a generic "Im" galaxy from the spectral synthesis models of Bruzual & Charlot (1993). The dotted line at $\mathcal{R} = 25.0$ is the limit reached in the present survey for what we have termed "robust" candidates.

Fig. 15—The spatial distribution of the "robust" $3 \leq z \leq 3.5$ galaxy candidates in the largest field in the present sample, Q0347−3819. There is no obvious tendency for the candidates to cluster around the QSO position, although the transverse size of the field is only 1.4 $h^{-1}$ Mpc for $q_0 = 0.05$ and 0.8 $h^{-1}$ Mpc for $q_0 = 0.5$˙

TABLE 1. Summary of Images

| QSO Field | Telescope | Filter | Int | FWHM | $SB_{lim}$[a] | $m_{lim}$[b] | $E_{B-V}$[c] |
|---|---|---|---|---|---|---|---|
| Q0347−3819 | CTIO 4m | $U_n$ | 14,000s | 1$''$.95 | 28.88 | 26.96 | 0.031 |
| | CTIO 4m | $G$ | 10,750s | 1$''$.96 | 29.21 | 27.29 | |
| | CTIO 4m | $\mathcal{R}$ | 4,000s | 1$''$.55 | 28.21 | 26.54 | |
| Q1244+1129 | WHT 4.2m | $U_n$ | 10,000s | 0$''$.86 | 28.38 | 27.35 | 0.037 |
| | WHT 4.2m | $G$ | 3,000s | 0$''$.85 | 28.24 | 27.22 | |
| | WHT 4.2m | $\mathcal{R}$ | 6,000s | 0$''$.89 | 28.56 | 27.49 | |
| Q1451+1223 | WHT 4.2m | $U_n$ | 9,200s | 0$''$.83 | 28.15 | 27.16 | 0.032 |
| | WHT 4.2m | $G$ | 3,600s | 0$''$.84 | 28.30 | 27.30 | |
| | MDM 2.4m | $\mathcal{R}$ | 9,000s | 0$''$.90 | 28.30 | 27.22 | |
| Q2233+1310 | WHT 4.2m | $U_n$ | 18,000s | 1$''$.34 | 28.77 | 27.23 | 0.084 |
| | MDM 2.4m | $G$ | 10,030s | 1$''$.01 | 28.44 | 27.24 | |
| | MDM 2.4m | $\mathcal{R}$ | 6,700s | 0$''$.89 | 28.33 | 27.26 | |
| | WHT 4.2m | $\mathcal{R}$ | 4,800s | 0$''$.90 | 28.21 | 27.13 | |

[a] $1\sigma$ surface brightness per square arc second, in AB magnitudes
[b] $3\sigma$ detection limit inside aperture the size of the seeing disk, in AB magnitudes
[c] Galactic extinction estimated as described in the text.

TABLE 2. QSO Photometry and LLS Absorber Redshifts

| QSO | $z_{em}$ | $z_{LLS}$ | $\mathcal{R}$ | $G$ | $U_n$ |
|---|---|---|---|---|---|
| Q0347−3819 | 3.222 | 3.025 | 17.73 | 18.24 | 20.59 |
| Q1244+1129 | 3.138 | 3.117 | 18.16 | 18.21 | 21.06 |
| Q1451+1223 | 3.247 | 3.171 | 19.16 | 19.14 | 22.85 |
| Q2233+1310 | 3.295 | 3.151 | 18.31 | 18.75 | 22.85 |

TABLE 3. Known Absorption Systems Toward Q0347−3819

| Redshift | Comments | Reference[a] |
|---|---|---|
| 1.4571 | Strong Mg II, Fe II | 1,2 |
| 1.5263 | Weak Mg II, Fe II | 1 |
| 2.3852 | Weak C IV, no low ions | 1 |
| 2.5706 | Weak C IV, no low ions | 1 |
| 2.6510 | Complex C IV, no low ions | 1 |
| 2.8103 | Moderate C IV, Si IV, no C II, Si II | 1 |
| 3.0244 | Damped Lyman $\alpha$, Moderate C IV, strong C II, Si II | 1,2 |

[a]References: 1) Williger et al. (1989) 2) Steidel (1990)

Table 4. Objects in the Field of Q0347−3819[a]

| ID No. | $\Delta\alpha$ ($''$) | $\Delta\delta$ ($''$) | $\Delta\theta$ ($''$) | $\mathcal{R}$ | $(G-\mathcal{R})$ | $(U_n-G)$ |
|---|---|---|---|---|---|---|
| 1 | −2.8 | 1.3 | 3.1 | 24.30 | 0.80 | −0.11 |
| 2 | −8.4 | −0.6 | 8.4 | 23.31 | 1.62 | 1.54 |
| 3 | 1.4 | 13.0 | 13.1 | 24.13 | 1.29 | −0.17 |
| 4 | 12.1 | −9.8 | 15.6 | 20.62 | 1.76 | 2.84 |
| 5 | −16.5 | −3.9 | 16.9 | 24.74 | 1.46 | −0.08 |
| 6 | 14.1 | 9.4 | 17.0 | 21.64 | 1.48 | 1.88 |
| 7 | −17.4 | 4.7 | 18.1 | 24.31 | 0.55 | 0.67 |
| 8 | 16.6 | −9.3 | 19.0 | 25.19 | 1.24 | 0.34 |
| 9 | 7.0 | 18.6 | 19.9 | 23.31 | 0.50 | 0.28 |
| 10 | 9.7 | −18.5 | 20.9 | 24.99 | 0.51 | −0.09 |
| 11 | −16.1 | 15.1 | 22.1 | 22.84 | 0.96 | 0.80 |
| 12 | 5.7 | 21.7 | 22.4 | 24.27 | 0.90 | 0.24 |
| 13 | −12.0 | 20.3 | 23.5 | 23.68 | 0.33 | 0.42 |
| 14 | −22.4 | −7.5 | 23.6 | 22.14 | 1.00 | 0.37 |
| 15 | 19.7 | −13.6 | 24.0 | 22.61 | 1.04 | 0.25 |
| 16 | −8.2 | 22.6 | 24.1 | 25.25 | 0.42 | 0.17 |
| 17 | 25.3 | 6.4 | 26.1 | 19.53 | 0.61 | 0.87 |
| 18 | 26.2 | 1.3 | 26.2 | 22.61 | 0.93 | 0.95 |
| 19 | 18.1 | 20.8 | 27.6 | 23.92 | 0.85 | 0.47 |
| 20 | 21.3 | −18.0 | 27.9 | 20.34 | 2.18 | 3.39 |
| 21 | −8.8 | 27.4 | 28.8 | 23.19 | 1.17 | 0.26 |
| 22 | −23.9 | 17.0 | 29.3 | 25.25 | 1.51 | −0.27 |
| 23 | 8.1 | −29.8 | 30.9 | 24.45 | 0.44 | 0.92 |
| 24 | −30.0 | −8.3 | 31.1 | 25.24 | 0.91 | 0.20 |
| 25 | −30.7 | 7.2 | 31.5 | 21.79 | 0.96 | 0.56 |
| 26 | −11.2 | 30.2 | 32.2 | 24.78 | 0.27 | 0.03 |
| 27 | 18.5 | −26.5 | 32.3 | 23.78 | 0.93 | 1.58 |
| 28 | −17.3 | −28.4 | 33.3 | 23.93 | 0.27 | 1.33 |
| 29 | −10.1 | −32.0 | 33.6 | 24.52 | 1.02 | 1.01 |
| 30 | 32.3 | −11.2 | 34.2 | 24.94 | 0.41 | 1.76 |
| 31 | −30.8 | 20.9 | 37.2 | 23.73 | 0.66 | 0.09 |
| 32 | −37.3 | −0.1 | 37.3 | 20.31 | 1.76 | 2.25 |
| 33 | −29.8 | −22.6 | 37.4 | 23.38 | 0.24 | 0.37 |
| 34 | 2.7 | 37.4 | 37.5 | 22.57 | 1.21 | 0.36 |
| 35 | 37.4 | −7.1 | 38.1 | 24.97 | 0.59 | 0.40 |
| 36 | 12.4 | −36.3 | 38.3 | 24.72 | 0.59 | 0.20 |
| 37 | 20.4 | −32.4 | 38.3 | 24.74 | 1.07 | 0.14 |
| 38 | −23.8 | −30.6 | 38.8 | 24.93 | 0.92 | 0.32 |
| 39 | 1.9 | −40.0 | 40.1 | 18.33 | 1.45 | 1.98 |
| 40 | 27.2 | 29.9 | 40.5 | 23.37 | 1.18 | 1.86 |
| 41 | 41.2 | 1.8 | 41.2 | 23.64 | 0.85 | 0.04 |
| 42 | 38.3 | 17.9 | 42.2 | 23.18 | 0.11 | 0.18 |
| 43 | −38.9 | −16.3 | 42.2 | 24.23 | 0.19 | 0.31 |
| 44 | −42.3 | −0.8 | 42.3 | 24.62 | 0.74 | 1.19 |
| 45 | −12.7 | −41.2 | 43.1 | 25.06 | 1.04 | 1.06 |
| 46 | −17.4 | −41.4 | 44.9 | 24.55 | 0.47 | 2.36 |
| 47 | −37.6 | −25.5 | 45.4 | 24.09 | 0.96 | 0.49 |

Table 4. (continued)

| ID No. | $\Delta\alpha$ ($''$ ) | $\Delta\delta$ ($''$ ) | $\Delta\theta$ ($''$ ) | $\mathcal{R}$ | $(G-\mathcal{R})$ | $(U_n-G)$ |
|---|---|---|---|---|---|---|
| 48 | 45.5 | 4.8 | 45.8 | 25.17 | 0.65 | 0.82 |
| 49 | 44.1 | 13.8 | 46.2 | 22.34 | 1.03 | 0.50 |
| 50 | 23.8 | −39.7 | 46.3 | 23.46 | 0.87 | 0.62 |
| 51 | 39.1 | −25.4 | 46.7 | 17.59 | 1.42 | 1.99 |
| 52 | −24.6 | −40.3 | 47.2 | 22.42 | 0.73 | 0.74 |
| 53 | 21.4 | −42.2 | 47.3 | 24.48 | 0.43 | 0.19 |
| 54 | 27.1 | 39.4 | 47.8 | 25.05 | 1.12 | 1.15 |
| 55 | 48.9 | −10.2 | 50.0 | 21.41 | 1.37 | 1.56 |
| 56 | 9.1 | 50.2 | 51.0 | 21.58 | 1.05 | 0.80 |
| 57 | −50.5 | −8.1 | 51.2 | 22.37 | 0.90 | 0.46 |
| 58 | −15.9 | −48.8 | 51.3 | 24.48 | 0.33 | 0.42 |
| 59 | 17.0 | 48.4 | 51.3 | 25.01 | 0.99 | 0.66 |
| 60 | 14.1 | 49.8 | 51.8 | 25.24 | 0.00 | 0.93 |
| 61 | 47.9 | 20.7 | 52.2 | 20.05 | 0.55 | 0.24 |
| 62 | 32.8 | −40.8 | 52.3 | 25.20 | 0.62 | 1.46 |
| 63 | 38.0 | 36.3 | 52.6 | 24.55 | 0.18 | 0.77 |
| 64 | 25.5 | 47.7 | 54.1 | 25.39 | 0.27 | 0.18 |
| 65 | −26.4 | 47.5 | 54.3 | 21.29 | 1.09 | 0.65 |
| 66 | 18.8 | −51.7 | 55.0 | 25.00 | 0.11 | 0.13 |
| 67 | −9.6 | −54.6 | 55.5 | 16.92 | 1.21 | 2.45 |
| 68 | −20.0 | 51.8 | 55.5 | 21.82 | 1.48 | 0.70 |
| 69 | −19.6 | −52.3 | 55.8 | 24.10 | 0.82 | 0.55 |
| 70 | −38.0 | 41.0 | 55.9 | 25.15 | 1.28 | 0.70 |
| 71 | −52.3 | −20.0 | 56.0 | 25.31 | 1.15 | 0.63 |
| 72 | −56.5 | 6.8 | 56.9 | 24.07 | 0.71 | 1.01 |
| 73 | −51.7 | 24.3 | 57.2 | 22.49 | 0.50 | 0.13 |
| 74 | 56.4 | 10.0 | 57.3 | 25.17 | 0.64 | 1.09 |
| 75 | 52.0 | 24.9 | 57.7 | 23.10 | 0.68 | 1.09 |
| 76 | 29.4 | −50.1 | 58.1 | 23.96 | 0.51 | 0.23 |
| 77 | −48.9 | 31.9 | 58.4 | 20.49 | 1.10 | 0.67 |
| 78 | 30.7 | 50.0 | 58.6 | 22.49 | 0.90 | 0.73 |
| 79 | −42.6 | −40.4 | 58.7 | 25.27 | 0.22 | 1.06 |
| 80 | −17.0 | −56.4 | 58.9 | 24.68 | 1.06 | 0.27 |
| 81 | 19.0 | −56.5 | 59.5 | 22.79 | 1.22 | 0.55 |
| N1 | −8.1 | −6.4 | 10.3 | 25.00 | 1.43 | >1.22 |
| N2 | 3.3 | −16.6 | 17.0 | 24.70 | 0.97 | >1.73 |
| N3 | 22.2 | −3.6 | 22.5 | 22.45 | 2.09 | >2.72 |
| N4 | 16.4 | 17.0 | 23.6 | 25.05 | 0.79 | >1.87 |
| N5 | −1.5 | 26.0 | 26.1 | 23.82 | 0.65 | >2.85 |
| N6 | −24.0 | 13.9 | 27.7 | 23.73 | 1.74 | >1.97 |
| N7 | −34.7 | −8.1 | 35.6 | 24.44 | 1.60 | >1.47 |
| N8 | −38.2 | 9.8 | 39.4 | 23.42 | 1.47 | >2.34 |
| N9 | −41.2 | 12.8 | 43.2 | 25.46 | 1.41 | >1.00 |
| N10 | 47.8 | −1.2 | 47.8 | 22.64 | 1.91 | >2.61 |
| N11 | −26.7 | 40.0 | 48.1 | 22.35 | 2.48 | >2.40 |
| N12 | −49.8 | 6.2 | 50.2 | 24.93 | 0.39 | >2.01 |

TABLE 4. (continued)

| ID No. | $\Delta\alpha$ ($''$ ) | $\Delta\delta$ ($''$ ) | $\Delta\theta$ ($''$ ) | $\mathcal{R}$ | $(G-\mathcal{R})$ | $(U_n-G)$ |
|---|---|---|---|---|---|---|
| N13 | −41.2 | 35.0 | 54.1 | 25.59 | 1.68 | >0.53 |
| N14 | 38.6 | −41.2 | 56.5 | 24.76 | 1.00 | >1.42 |
| N15 | −50.8 | −27.4 | 57.7 | 25.41 | 0.73 | >1.41 |

[a]Only objects within 60$''$ of the QSO are included

TABLE 5. Absorption Lines in the Spectrum of Q1244+1129

| No. | $\lambda_{obs}$ | $\sigma(\lambda)$ | $W_{obs}$ | $\sigma(W)$ | S/N | ID | $z_{abs}$ |
|---|---|---|---|---|---|---|---|
| 1 | 4418.61 | 0.15 | 27.74 | 0.31 | 25.1 | HI(1215) [a] | 2.6347 |
| 2 | 4978.02 | 0.11 | 35.96 | 0.24 | 34.7 | HI(1215)[a] | 3.0950 |
| 3 | 5037.85 | 0.13 | 2.30 | 0.07 | 65.4 | | |
| 4 | 5063.94 | 0.08 | 9.35 | 0.08 | 70.6 | | |
| 5 | 5137.68 | 0.38 | 0.75 | 0.08 | 47.9 | | |
| 6 | 5164.02 | 0.08 | 5.14 | 0.08 | 52.0 | SiII(1260) | 3.0971 |
| 7 | 5335.75 | 0.11 | 4.18 | 0.12 | 32.9 | OI(1302) | 3.0976 |
| 8 | 5344.65 | 0.11 | 3.31 | 0.12 | 32.9 | SiII(1304) | 3.0975 |
| 9 | 5468.87 | 0.12 | 7.05 | 0.15 | 32.8 | CII(1334) | 3.0980 |
| 10 | 5497.65 | 0.32 | 2.03 | 0.23 | 15.8 | AlIII(1862) | 1.9513 |
| 11 | 5551.84 | 0.49 | 1.93 | 0.26 | 15.8 | SiII(1526) | 2.6364 |
| 12 | 5633.05 | 0.43 | 5.84 | 0.33 | 18.3 | CIV(1549) | 2.6361 |
| 13 | 5710.29 | 0.18 | 2.72 | 0.17 | 22.0 | SiIV(1393) | 3.0970 |
| 14 | 5747.99 | 0.24 | 1.41 | 0.14 | 24.7 | SiIV(1402) | 3.0976 |
| 15 | 5848.26 | 0.47 | 0.79 | 0.13 | 27.2 | FeII(1608) | 2.6359 |
| 16 | 5899.02 | 0.37 | 0.87 | 0.13 | 25.6 | | |
| 17 | 6075.50 | 0.18 | 2.87 | 0.14 | 29.8 | AlII(1670) | 2.6363 |
| 18 | 6217.59 | 0.29 | 0.94 | 0.11 | 30.4 | | |
| 19 | 6255.63 | 0.11 | 4.21 | 0.12 | 33.2 | SiII(1526) | 3.0974 |
| 20 | 6343.70 | 0.11 | 4.33 | 0.13 | 30.9 | CIV(1548) | 3.0975 |
| 21 | 6354.99 | 0.19 | 3.14 | 0.14 | 30.5 | CIV(1550) | 3.0980 |
| 22 | 6384.89 | 0.30 | 1.10 | 0.11 | 31.7 | MgII(2796) | 1.2833 |
| 23 | 6401.27 | 0.43 | 0.76 | 0.12 | 31.2 | MgII(2803) | 1.2833 |
| 24 | 6447.15 | 0.40 | 0.42 | 0.08 | 37.1 | | |
| 25 | 6590.79 | 0.16 | 2.12 | 0.11 | 32.4 | FeII(1608) | 3.0976 |
| 26 | 6846.31 | 0.11 | 4.39 | 0.13 | 28.9 | AlII(1670) | 3.0976 |
| 27 | 6917.79 | 0.35 | 1.86 | 0.18 | 22.9 | FeII(2344) | 1.9510 |
| 28 | 7005.88 | 0.18 | 1.20 | 0.09 | 38.1 | FeII(2374) | 1.9514 |
| 29 | 7031.15 | 0.14 | 2.53 | 0.10 | 42.0 | FeII(2382) | 1.9508 |
| 30 | 7137.96 | 0.30 | 0.77 | 0.09 | 35.9 | | |
| 31 | 7407.59 | 0.25 | 0.88 | 0.09 | 38.1 | SiII(1808) | 3.0971 |
| 32 | 7603.21 | 0.27 | 5.75 | 0.23 | 23.6 | b | |
| 33 | 7629.12 | 0.27 | 3.42 | 0.20 | 22.5 | b | |
| 34 | 7672.96 | 0.25 | 2.19 | 0.19 | 19.2 | FeII(2600) | 1.9509 |
| 35 | 8252.15 | 0.13 | 5.72 | 0.20 | 21.0 | MgII(2796) | 1.9510 |
| 36 | 8273.33 | 0.10 | 4.65 | 0.17 | 20.1 | MgII(2803) | 1.9510 |
| 37 | 8419.44 | 0.34 | 1.89 | 0.20 | 19.5 | MgI(2852) | 1.9511 |

[a]Feature is associated with identified metal line system

[b]These lines are probably spurious and are due to imperfect removal of the atmospheric A–band

TABLE 6. Objects in the Field of Q1244+1129

| ID No. | $\Delta\alpha$ ($''$) | $\Delta\delta$ ($''$) | $\Delta\theta$ ($''$) | $\mathcal{R}$ | $(G-\mathcal{R})$ | $(U_n-G)$ |
|---|---|---|---|---|---|---|
| 1 | 3.7 | 3.7 | 5.2 | 25.36 | 0.12 | 0.86 |
| 2 | −2.1 | 6.5 | 6.9 | 25.25 | −0.31 | 0.15 |
| 3 | −7.0 | −1.0 | 7.1 | 23.62 | 0.04 | 0.94 |
| 4 | 7.6 | 3.8 | 8.5 | 25.45 | 0.02 | 0.70 |
| 5 | 8.4 | −5.6 | 10.1 | 23.95 | 0.61 | 0.36 |
| 6 | 10.2 | −0.5 | 10.2 | 25.48 | 0.76 | 0.26 |
| 7 | 5.5 | 10.4 | 11.7 | 25.41 | 0.44 | 0.62 |
| 8 | 2.2 | −12.5 | 12.7 | 24.59 | 0.08 | 2.66 |
| 9 | −10.0 | −8.1 | 12.8 | 24.43 | 0.96 | 0.82 |
| 10 | −0.7 | −13.4 | 13.4 | 23.37 | 0.63 | 0.29 |
| 11 | 13.9 | 2.8 | 14.2 | 23.71 | 0.66 | 0.10 |
| 12 | −8.9 | 13.1 | 15.8 | 24.35 | 0.77 | 2.14 |
| 13 | −14.2 | −8.3 | 16.5 | 25.23 | 0.35 | 0.42 |
| 14 | 6.4 | −15.7 | 17.0 | 24.83 | 0.38 | 0.71 |
| 15 | −1.4 | 17.9 | 17.9 | 20.82 | 0.65 | 0.62 |
| 16 | −15.6 | 9.4 | 18.3 | 25.12 | 0.51 | 0.10 |
| 17 | −18.3 | 2.6 | 18.5 | 23.41 | 0.89 | 0.66 |
| 18 | −7.7 | −16.8 | 18.5 | 24.52 | 1.28 | 0.11 |
| 19 | −8.9 | −18.7 | 20.7 | 22.85 | 0.74 | 0.50 |
| 20 | 17.6 | −12.3 | 21.4 | 25.31 | 0.18 | 0.33 |
| 21 | 14.4 | 17.4 | 22.6 | 22.46 | 0.47 | 1.52 |
| 22 | −3.2 | 23.3 | 23.6 | 23.32 | 1.65 | 1.74 |
| 23 | −10.1 | 21.6 | 23.8 | 23.57 | 0.73 | 0.11 |
| 24 | 10.7 | −21.9 | 24.4 | 20.91 | 0.00 | 0.85 |
| 25 | −15.0 | 19.8 | 24.9 | 25.24 | 0.62 | 1.39 |
| 26 | 19.9 | −16.4 | 25.8 | 22.91 | 1.52 | 2.40 |
| 27 | −24.7 | −7.6 | 25.8 | 24.61 | 0.17 | 0.12 |
| 28 | −9.8 | −24.0 | 26.0 | 23.47 | 2.36 | 0.76 |
| 29 | −4.5 | 25.9 | 26.2 | 21.33 | 1.13 | 0.74 |
| 30 | 23.1 | −16.2 | 28.2 | 22.36 | 0.58 | 0.63 |
| 31 | 14.8 | −24.3 | 28.5 | 24.92 | 1.60 | −0.51 |
| 32 | 26.8 | −11.3 | 29.0 | 25.17 | 0.54 | 0.61 |
| 33 | −30.0 | 5.7 | 30.5 | 24.62 | 1.29 | 0.87 |
| 34 | −5.2 | −30.3 | 30.7 | 23.98 | 0.56 | 0.25 |
| 35 | −19.7 | −24.2 | 31.2 | 24.35 | 0.56 | 1.24 |
| 36 | 31.4 | 0.5 | 31.4 | 23.43 | 0.77 | 0.53 |
| 37 | −29.9 | −11.0 | 31.8 | 23.93 | 0.60 | 1.62 |
| 38 | 29.1 | 13.2 | 31.9 | 23.82 | 0.77 | 1.50 |
| 39 | −7.2 | 31.4 | 32.3 | 24.39 | 0.02 | 0.04 |
| 40 | 27.8 | −16.8 | 32.5 | 24.43 | 0.38 | 2.02 |
| 41 | 3.2 | −32.4 | 32.5 | 25.24 | 1.04 | 0.68 |
| 42 | 31.6 | −8.7 | 32.8 | 25.43 | −0.02 | 1.09 |
| 43 | −26.3 | −20.3 | 33.2 | 24.79 | 0.54 | −0.08 |
| 44 | −31.3 | 11.0 | 33.2 | 25.06 | 0.70 | 0.99 |
| 45 | 33.0 | −4.7 | 33.3 | 23.86 | 0.11 | 0.56 |
| 46 | −24.5 | −23.5 | 34.0 | 23.82 | 0.57 | 0.11 |
| 47 | 17.0 | 29.7 | 34.2 | 25.00 | 0.08 | 0.40 |

Table 6. (continued)

| ID No. | $\Delta\alpha$ ($''$) | $\Delta\delta$ ($''$) | $\Delta\theta$ ($''$) | $\mathcal{R}$ | $(G-\mathcal{R})$ | $(U_n-G)$ |
|---|---|---|---|---|---|---|
| 48 | 32.7 | 10.0 | 34.2 | 25.52 | 0.02 | 0.32 |
| 49 | 10.9 | −34.1 | 35.8 | 25.32 | −0.16 | 0.93 |
| 50 | 24.9 | −26.1 | 36.1 | 25.00 | 0.32 | 0.54 |
| 51 | 14.7 | −33.8 | 36.9 | 24.17 | 0.36 | 0.04 |
| 52 | 28.1 | 24.8 | 37.4 | 24.40 | 0.35 | 0.16 |
| 53 | 8.8 | 36.4 | 37.5 | 25.33 | 0.16 | 0.84 |
| 54 | 29.5 | 23.9 | 38.0 | 23.09 | 0.81 | 0.67 |
| 55 | −1.2 | 38.1 | 38.1 | 25.27 | 0.37 | −0.24 |
| 56 | −33.5 | 18.7 | 38.3 | 24.04 | 1.09 | 0.44 |
| 57 | −38.3 | −5.2 | 38.7 | 24.66 | 1.17 | 0.91 |
| 58 | −36.3 | 14.4 | 39.0 | 22.29 | 0.74 | 0.17 |
| 59 | 36.7 | 14.8 | 39.6 | 22.07 | 0.60 | 0.51 |
| 60 | 1.9 | −40.3 | 40.3 | 21.84 | 0.43 | 0.58 |
| 61 | 12.3 | −38.8 | 40.7 | 20.15 | 0.79 | 1.19 |
| 62 | −35.4 | 21.0 | 41.1 | 25.11 | 0.65 | −0.08 |
| 63 | 5.7 | −40.8 | 41.2 | 23.27 | 0.57 | 0.12 |
| 64 | −10.9 | −39.8 | 41.3 | 24.80 | −0.35 | 0.21 |
| 65 | 37.9 | −17.6 | 41.8 | 23.95 | 0.76 | 0.49 |
| 66 | 40.6 | −11.3 | 42.2 | 25.34 | 0.00 | 0.60 |
| 67 | −22.6 | −36.2 | 42.7 | 24.58 | 0.55 | 0.83 |
| 68 | −20.2 | −38.1 | 43.1 | 23.31 | 1.07 | 0.93 |
| 69 | 0.2 | −43.7 | 43.7 | 21.86 | 0.88 | 1.07 |
| 70 | −22.8 | 38.2 | 44.5 | 22.81 | 0.54 | 0.34 |
| 71 | −21.8 | −39.3 | 44.9 | 24.06 | 0.49 | 0.06 |
| 72 | 43.5 | 13.4 | 45.5 | 22.13 | 0.79 | 0.36 |
| 73 | −41.5 | 20.2 | 46.2 | 25.10 | −0.21 | 0.12 |
| 74 | −18.9 | −42.4 | 46.4 | 24.39 | 0.07 | 0.08 |
| 75 | 33.9 | −32.8 | 47.1 | 24.88 | 0.18 | 1.11 |
| 76 | 42.3 | 23.0 | 48.1 | 25.04 | 0.62 | 0.64 |
| 77 | 23.6 | −42.6 | 48.7 | 24.11 | 0.03 | 0.16 |
| 78 | 33.2 | −36.5 | 49.4 | 24.52 | 1.10 | 1.57 |
| 79 | 43.2 | 24.3 | 49.6 | 24.29 | 0.54 | 0.20 |
| 80 | −25.1 | −42.8 | 49.6 | 25.25 | 0.63 | 0.84 |
| 81 | −38.5 | −32.0 | 50.0 | 23.01 | 1.10 | 0.93 |
| 82 | −38.3 | −37.3 | 53.5 | 24.27 | 0.29 | 0.16 |
| 83 | 42.1 | −39.5 | 57.7 | 24.25 | −0.08 | 0.01 |
| 84 | 45.6 | 35.7 | 57.9 | 22.82 | 0.63 | 0.49 |
| N1 | 17.0 | −8.2 | 18.9 | 25.78 | 1.13 | >0.94 |
| N2 | −18.8 | 21.3 | 28.4 | 24.68 | 1.71 | >1.35 |
| N3 | 22.3 | 19.0 | 29.3 | 24.05 | 1.35 | >2.18 |
| N4 | −38.3 | −6.8 | 38.9 | 24.31 | 1.99 | >1.33 |
| N5 | 33.1 | −22.5 | 40.0 | 24.90 | 0.95 | >1.68 |
| N6 | 40.6 | 5.0 | 40.9 | 24.66 | 0.69 | >2.45 |
| N7 | −29.1 | 35.5 | 45.9 | 22.30 | 1.59 | >3.13 |

Table 7. Objects in the Field of Q1451+1223

| ID No. | $\Delta\alpha$ ($''$) | $\Delta\delta$ ($''$) | $\Delta\theta$ ($''$) | $\mathcal{R}$ | $(G-\mathcal{R})$ | $(U_n-G)$ |
|---|---|---|---|---|---|---|
| 1 | −6.2 | −5.3 | 8.2 | 23.79 | 0.52 | 0.21 |
| 2 | −8.6 | 1.5 | 8.7 | 21.47 | 0.29 | 1.19 |
| 3 | 8.5 | −5.7 | 10.3 | 21.74 | 0.63 | 0.73 |
| 4 | 1.4 | 10.4 | 10.5 | 23.15 | 0.81 | 0.22 |
| 5 | 5.3 | 11.9 | 13.0 | 25.05 | −0.12 | 1.80 |
| 6 | −13.5 | 5.4 | 14.6 | 25.38 | 0.65 | 0.79 |
| 7 | 11.5 | 11.0 | 15.9 | 23.82 | 0.24 | 0.20 |
| 8 | 18.0 | 1.3 | 18.1 | 23.71 | 1.15 | 1.29 |
| 9 | 1.1 | 21.5 | 21.5 | 25.06 | 0.21 | 0.58 |
| 10 | 1.8 | 23.4 | 23.4 | 25.12 | 1.14 | 0.41 |
| 11 | 24.8 | −7.9 | 26.1 | 24.12 | 0.27 | 1.85 |
| 12 | 10.4 | 26.0 | 28.0 | 25.29 | 0.01 | 1.75 |
| 13 | −14.6 | 23.8 | 28.0 | 25.34 | 0.56 | −0.11 |
| 14 | 26.8 | −8.3 | 28.0 | 25.51 | 0.14 | 1.75 |
| 15 | −16.2 | 23.0 | 28.1 | 24.99 | 0.26 | 0.43 |
| 16 | −2.1 | 28.8 | 28.9 | 20.69 | 0.40 | 1.37 |
| 17 | 30.2 | −8.9 | 31.5 | 21.95 | 0.59 | 1.09 |
| 18 | −28.0 | 16.2 | 32.3 | 24.70 | −0.01 | 0.79 |
| 19 | 12.9 | 30.8 | 33.4 | 23.32 | 0.99 | 2.11 |
| 20 | −33.6 | 8.1 | 34.6 | 22.54 | 1.31 | 0.89 |
| 21 | 0.4 | 38.0 | 38.0 | 22.70 | 1.75 | 0.94 |
| 22 | 10.6 | 37.3 | 38.7 | 24.51 | 0.67 | 0.78 |
| 23 | 23.5 | 30.9 | 38.9 | 22.83 | 0.24 | 0.92 |
| 24 | 31.0 | 23.9 | 39.1 | 24.92 | 0.21 | 0.46 |
| 25 | −5.7 | 39.3 | 39.7 | 24.76 | −0.17 | 0.88 |
| 26 | −25.4 | 30.8 | 39.9 | 23.91 | 0.37 | 0.21 |
| 27 | −12.9 | 37.8 | 40.0 | 21.75 | 1.45 | 3.19 |
| 28 | −31.3 | 25.7 | 40.5 | 23.21 | 0.83 | 0.69 |
| 29 | −10.1 | 39.4 | 40.7 | 24.03 | 0.32 | 0.32 |
| 30 | −19.3 | 37.1 | 41.8 | 25.04 | 0.00 | 0.15 |
| 31 | −39.8 | 16.3 | 43.0 | 25.32 | 0.19 | 1.13 |
| 32 | −8.7 | 42.4 | 43.3 | 24.58 | −0.18 | 0.37 |
| 33 | −13.8 | 41.7 | 43.9 | 23.70 | 0.18 | 0.64 |
| 34 | −33.8 | 33.0 | 47.2 | 23.95 | 1.96 | 0.36 |
| 35 | 26.1 | 40.8 | 48.5 | 25.49 | 0.46 | 1.10 |
| 36 | −39.1 | 29.0 | 48.7 | 24.27 | 1.24 | 0.14 |
| 37 | −41.9 | 29.5 | 51.3 | 23.61 | 0.49 | 1.44 |
| 38 | −12.5 | 51.9 | 53.3 | 24.07 | −0.10 | 0.49 |
| 39 | −42.3 | 35.5 | 55.2 | 22.05 | 0.86 | 0.70 |
| 40 | −26.8 | 49.9 | 56.7 | 24.49 | 0.13 | 0.57 |
| 41 | −17.3 | 54.3 | 57.0 | 25.35 | 0.42 | 0.26 |
| 42 | −44.5 | 35.9 | 57.1 | 23.91 | 0.85 | 0.82 |
| 43 | −13.2 | 57.4 | 58.9 | 25.45 | 0.99 | −0.50 |
| 44 | −5.0 | 62.0 | 62.2 | 24.36 | 0.22 | 0.76 |
| 45 | 9.3 | 63.6 | 64.3 | 22.02 | 0.67 | 0.51 |
| 46 | 9.3 | 66.4 | 67.1 | 25.07 | 1.23 | 1.47 |
| 47 | 9.1 | 67.8 | 68.4 | 23.78 | 1.56 | 1.75 |

Table 7. (continued)

| ID No. | $\Delta\alpha$ ($''$) | $\Delta\delta$ ($''$) | $\Delta\theta$ ($''$) | $\mathcal{R}$ | $(G-\mathcal{R})$ | $(U_n-G)$ |
|---|---|---|---|---|---|---|
| 48 | −10.8 | 67.7 | 68.6 | 23.52 | 0.45 | 0.37 |
| 49 | −26.3 | 63.5 | 68.8 | 20.88 | 0.77 | 1.10 |
| 50 | −42.5 | 55.8 | 70.1 | 25.03 | 1.03 | 0.09 |
| 51 | 30.9 | 63.7 | 70.8 | 23.27 | 0.07 | 1.26 |
| 52 | −39.3 | 61.2 | 72.7 | 22.60 | 0.25 | 0.19 |
| 53 | 34.7 | 69.0 | 77.2 | 24.32 | 0.27 | 1.97 |
| 54 | 30.7 | 72.3 | 78.6 | 20.88 | 1.94 | 2.18 |
| 55 | −27.1 | 74.5 | 79.3 | 23.79 | 0.97 | 0.91 |
| 56 | −36.3 | 74.2 | 82.6 | 25.27 | −0.03 | 0.44 |
| 57 | 35.8 | 75.1 | 83.2 | 24.30 | 0.90 | 1.65 |
| 58 | −40.4 | 76.9 | 86.8 | 24.89 | 0.57 | 1.22 |
| 59 | 15.6 | 86.7 | 88.1 | 24.76 | 0.43 | 0.54 |
| 60 | −5.5 | 89.8 | 90.0 | 16.25 | 0.18 | 1.11 |
| 61 | −17.4 | 89.4 | 91.1 | 25.08 | 0.44 | 0.88 |
| 62 | −22.1 | 88.9 | 91.6 | 25.32 | 0.72 | 0.59 |
| 63 | −28.0 | 92.1 | 96.2 | 23.30 | 0.78 | 0.67 |
| 64 | 21.0 | 94.0 | 96.4 | 24.98 | 1.31 | 0.10 |
| 65 | 18.0 | 95.8 | 97.5 | 21.41 | 1.14 | 0.73 |
| 66 | −31.9 | 94.1 | 99.4 | 24.28 | 0.07 | 1.31 |
| N1 | 20.5 | 27.2 | 34.1 | 24.45 | 0.05 | >2.96 |
| N2 | −34.2 | 14.1 | 37.0 | 22.35 | 2.19 | >2.17 |
| N3 | −7.8 | 48.2 | 48.8 | 25.19 | 0.69 | >1.63 |
| N4 | −34.0 | 37.7 | 50.7 | 24.20 | 1.68 | >1.52 |
| N5 | −31.9 | 41.8 | 52.6 | 25.13 | 0.11 | >2.09 |
| N6 | 21.9 | 48.7 | 53.4 | 25.51 | 0.62 | >1.62 |
| N7 | 33.5 | 51.6 | 61.5 | 24.04 | 1.22 | >2.19 |
| N8 | −7.3 | 62.0 | 62.5 | 24.77 | 0.28 | >2.34 |
| N9 | −0.3 | 72.7 | 72.7 | 25.54 | 0.64 | >1.40 |
| N10 | 21.7 | 90.1 | 92.7 | 24.85 | 1.46 | >1.24 |

TABLE 8. Absorption Lines in the Spectrum of Q2233+1310

| No. | $\lambda_{\rm obs}$ | $\sigma(\lambda)$ | $W_{\rm obs}$ | $\sigma(W)$ | S/N | ID | $z_{\rm abs}$ |
|---|---|---|---|---|---|---|---|
| 1 | 4318.53 | 0.14 | 20.23 | 0.34 | 20.7 | HI(1215)[a] | 2.5524 |
| 2 | 5046.20 | 0.15 | 23.09 | 0.31 | 25.9 | HI(1215)[a] | 3.1510 |
| 3 | 5230.64 | 0.11 | 3.73 | 0.10 | 50.4 | SiII(1260) | 3.1499 |
| 4 | 5403.38 | 0.14 | 2.30 | 0.11 | 30.3 | OI(1302) | 3.1495 |
| 5 | 5413.24 | 0.14 | 1.93 | 0.08 | 41.7 | SiII(1304) | 3.1501 |
| 6 | 5425.00 | 0.20 | 1.04 | 0.08 | 43.8 | SiII(1526) | 2.5534 |
| 7 | 5502.58 | 0.13 | 1.73 | 0.08 | 42.6 | CIV(1548) | 2.5542 |
| 8 | 5511.39 | 0.18 | 1.27 | 0.08 | 40.4 | CIV(1550) | 2.5540 |
| 9 | 5538.33 | 0.11 | 4.44 | 0.10 | 41.5 | CII(1334) | 3.1500 |
| 10 | 5666.53 | 0.15 | 0.50 | 0.05 | 46.2 | CIV(1548) | 2.6601 |
| 11 | 5675.51 | 0.31 | 0.43 | 0.06 | 44.2 | CIV(1550) | 2.6598 |
| 12 | 5785.14 | 0.15 | 1.48 | 0.08 | 42.0 | SiIV(1393) | 3.1507 |
| 13 | 5822.48 | 0.25 | 0.65 | 0.08 | 36.4 | SiIV(1402) | 3.1507 |
| 14 | 5938.87 | 0.23 | 0.62 | 0.08 | 32.2 | AlII(1670) | 2.5545 |
| 15 | 6336.13 | 0.10 | 2.12 | 0.07 | 49.2 | SiII(1526) | 3.1502 |
| 16 | 6427.65 | 0.16 | 2.01 | 0.08 | 45.9 | CIV(1548) | 3.1517 |
| 17 | 6438.60 | 0.18 | 0.77 | 0.06 | 45.0 | CIV(1550) | 3.1519 |
| 18 | 6934.78 | 0.22 | 2.30 | 0.10 | 60.3 | AlII(1670) | 3.1505 |

[a]Feature is associated with identified metal line system

Table 9. Objects in the Field of Q2233+1310

| ID No. | $\Delta\alpha$ ($''$) | $\Delta\delta$ ($''$) | $\Delta\theta$ ($''$) | $\mathcal{R}$ | $(G-\mathcal{R})$ | $(U_n-G)$ |
|---|---|---|---|---|---|---|
| 1 | 0.4 | −7.8 | 7.8 | 25.15 | 0.64 | 1.33 |
| 2 | −9.6 | 2.2 | 9.9 | 22.20 | 1.49 | 1.00 |
| 3 | 10.5 | 1.9 | 10.7 | 20.92 | 1.97 | 3.12 |
| 4 | 10.7 | −8.1 | 13.4 | 24.32 | 1.36 | 1.62 |
| 5 | −9.8 | 9.1 | 13.4 | 24.43 | 0.75 | 1.18 |
| 6 | 10.0 | 9.0 | 13.4 | 24.87 | 0.41 | 1.02 |
| 7 | 5.8 | −12.7 | 13.9 | 23.49 | 0.26 | 0.82 |
| 8 | 8.4 | 13.2 | 15.6 | 24.88 | 0.10 | 1.41 |
| 9 | 11.0 | −12.6 | 16.7 | 24.87 | 0.15 | 1.80 |
| 10 | 17.8 | −2.0 | 18.0 | 19.20 | 0.49 | 1.07 |
| 11 | 0.0 | 19.8 | 19.8 | 23.59 | 0.14 | 0.52 |
| 12 | −9.3 | −17.8 | 20.1 | 24.49 | −0.01 | 1.39 |
| 13 | −19.4 | −6.3 | 20.4 | 22.58 | 1.17 | 1.15 |
| 14 | 10.5 | −18.2 | 21.0 | 24.04 | 0.50 | 0.84 |
| 15 | −13.8 | 17.7 | 22.4 | 25.11 | 0.59 | 0.27 |
| 16 | 16.4 | 16.0 | 22.9 | 23.77 | 0.23 | 0.58 |
| 17 | −2.9 | 23.2 | 23.4 | 20.78 | 1.88 | 3.02 |
| 18 | 7.5 | 22.8 | 24.0 | 20.53 | 2.37 | 2.80 |
| 19 | 18.2 | −16.1 | 24.2 | 24.63 | 1.11 | 0.46 |
| 20 | −24.4 | 4.7 | 24.9 | 24.14 | 1.25 | −0.06 |
| 21 | 4.3 | 26.3 | 26.7 | 23.90 | 0.79 | 1.37 |
| 22 | 25.1 | −10.5 | 27.2 | 24.22 | 0.84 | 1.58 |
| 23 | 12.3 | 24.4 | 27.3 | 24.20 | 0.93 | 0.44 |
| 24 | −25.9 | 8.9 | 27.4 | 21.79 | 1.33 | 1.20 |
| 25 | −23.3 | 16.0 | 28.3 | 24.12 | 0.51 | 0.42 |
| 26 | 26.9 | −12.4 | 29.6 | 25.02 | 0.43 | 0.66 |
| 27 | −28.1 | −11.7 | 30.5 | 23.06 | 0.82 | 0.40 |
| 28 | 21.2 | −22.4 | 30.8 | 23.86 | −0.13 | 1.03 |
| 29 | 29.1 | −10.4 | 30.9 | 22.88 | 0.53 | 0.47 |
| 30 | −10.8 | −29.0 | 30.9 | 22.88 | 1.64 | 0.57 |
| 31 | −16.1 | −26.7 | 31.2 | 25.14 | 0.12 | 0.79 |
| 32 | 31.0 | 5.7 | 31.5 | 25.33 | 0.83 | 1.07 |
| 33 | 7.2 | −30.8 | 31.7 | 25.08 | 0.94 | 1.14 |
| 34 | −28.6 | 14.1 | 31.9 | 23.91 | 0.53 | 1.70 |
| 35 | 32.4 | −6.6 | 33.0 | 24.02 | 0.08 | 1.58 |
| 36 | 5.1 | 33.3 | 33.6 | 24.67 | 1.23 | 1.05 |
| 37 | −33.9 | −4.0 | 34.2 | 24.25 | 0.60 | 0.88 |
| 38 | −29.3 | −18.2 | 34.5 | 22.98 | 1.31 | 2.55 |
| 39 | −14.6 | −31.2 | 34.5 | 24.53 | 0.29 | 1.28 |
| 40 | −7.1 | −34.2 | 34.9 | 24.98 | 0.36 | 1.64 |
| 41 | −27.0 | −22.4 | 35.0 | 20.98 | 1.21 | 0.89 |
| 42 | 29.4 | −19.0 | 35.0 | 24.60 | 0.79 | 0.37 |
| 43 | 33.8 | 13.6 | 36.4 | 17.70 | 1.47 | 2.96 |
| 44 | 16.5 | 33.1 | 37.0 | 25.09 | 0.54 | 0.62 |
| 45 | −20.6 | −31.1 | 37.3 | 24.50 | 0.89 | 0.48 |
| 46 | −30.0 | 22.7 | 37.6 | 23.82 | 0.42 | 0.40 |
| 47 | 20.4 | 33.4 | 39.1 | 23.01 | 0.12 | 1.70 |

Table 9. (continued)

| ID No. | $\Delta\alpha$ ('') | $\Delta\delta$ ('') | $\Delta\theta$ ('') | $\mathcal{R}$ | $(G-\mathcal{R})$ | $(U_n-G)$ |
|---|---|---|---|---|---|---|
| 48 | −34.2 | −22.5 | 40.9 | 24.32 | 0.69 | 0.79 |
| 49 | −17.6 | −37.9 | 41.8 | 24.70 | 1.31 | 0.22 |
| 50 | 17.3 | −39.4 | 43.0 | 23.71 | 1.18 | 0.64 |
| 51 | −32.2 | −30.7 | 44.5 | 24.12 | 0.21 | 0.71 |
| 52 | −25.6 | −36.5 | 44.6 | 25.31 | −0.23 | 0.80 |
| 53 | −28.4 | 34.6 | 44.7 | 24.35 | 1.41 | −2.45 |
| 54 | 29.0 | 34.2 | 44.9 | 23.65 | 0.48 | 1.64 |
| 55 | 44.1 | −12.8 | 45.9 | 17.57 | 2.02 | 3.05 |
| 56 | 34.7 | 30.0 | 45.9 | 19.68 | 0.71 | 1.38 |
| 57 | 27.7 | −36.6 | 45.9 | 23.16 | 0.91 | 1.04 |
| 58 | −20.6 | −42.0 | 46.8 | 24.48 | 0.35 | 1.14 |
| 59 | 42.2 | 26.0 | 49.6 | 22.64 | 0.97 | 1.13 |
| 60 | 30.9 | −39.0 | 49.7 | 24.33 | 0.15 | 0.45 |
| 61 | 40.6 | −30.5 | 50.8 | 17.82 | 1.49 | 3.05 |
| 62 | 35.3 | −36.7 | 50.9 | 24.19 | 0.24 | 0.60 |
| 63 | 44.3 | −28.0 | 52.4 | 24.66 | 0.74 | 0.95 |
| N1 | 1.0 | −2.7 | 2.9 | 25.02 | 0.15 | >2.38 |
| N2 | −22.0 | −9.5 | 24.0 | 21.02 | 2.29 | >3.90 |
| N3 | 34.9 | 3.3 | 35.1 | 25.35 | 0.49 | >1.78 |
| N4 | −30.1 | 19.2 | 35.6 | 25.07 | 0.55 | >2.08 |
| N5 | 28.5 | 27.2 | 39.4 | 24.82 | 1.37 | >1.45 |
| N6 | 8.2 | −39.1 | 40.0 | 25.52 | 0.85 | >1.44 |
| N7 | −30.7 | −33.5 | 45.4 | 22.17 | 1.59 | >3.78 |
| N8 | 24.6 | −41.8 | 48.5 | 24.68 | 0.96 | >1.92 |

TABLE 10. Surface Density of High-$z$ Candidates

| QSO Field | Area Sampled (arcmin$^2$) | Number | Surface Density (arcmin$^{-2}$) |
|---|---|---|---|
| Q0000−2620 | 15.8 | 7 | $0.4 \pm 0.2$ |
| Q0347−3819 | 14.6 | 8 | $0.6 \pm 0.2$ |
| Q1244+1129 | 2.1 | 2 | $1.0 \pm 0.7$ |
| Q1451+1223 | 2.2 | 3 | $1.4 \pm 0.8$ |
| Q2233+1310 | 1.8 | 2 | $1.1 \pm 0.8$ |